\documentclass[12pt,a4paper]{article}
\usepackage[utf8]{inputenc}

\usepackage{framed}
\usepackage{mdframed}

\usepackage{fullpage}
\usepackage[T1]{fontenc}

\usepackage{epsf}
\usepackage{graphicx}
\usepackage{verbatim}
\usepackage{color}	
\usepackage{bbold}
\usepackage{hyperref}
\usepackage[dvipsnames]{xcolor}

\usepackage[most]{tcolorbox}
\usepackage{empheq}
\newtcbox{\mymath}[1][]{%
    nobeforeafter, math upper, tcbox raise base,
    enhanced, colframe=blue!30!black,
    colback=blue!30, boxrule=1pt,
    #1}

\usepackage[british]{babel}
\usepackage{verbatim}
\usepackage[T1]{fontenc}
\usepackage{lmodern}
\usepackage{lipsum}
\usepackage{booktabs}
\usepackage{caption}
\usepackage{cite}
\usepackage{soul,color}
\usepackage[toc,page]{appendix}

\usepackage{ifpdf}
\usepackage{cancel}

\usepackage{amsthm, amssymb}
\usepackage{pifont}
\usepackage{bm}
\usepackage{mathtools}
\usepackage{amstext}
\usepackage{braket}
\usepackage{multirow}

\usepackage[normalem]{ulem}
\usepackage{xcolor}
\usepackage{cancel}

\usepackage{lmodern} 
\usepackage{diagbox}

\newcommand\redsout{\bgroup\markoverwith{\textcolor{red}{\rule[0.5ex]{2pt}{0.4pt}}}\ULon}

\begin{document}
\vspace{5mm}
\vspace{0.5cm}

\def\be{\begin{eqnarray}}
\def\ee{\end{eqnarray}}

\def\ba{\begin{aligned}}
\def\ea{\end{aligned}}

\def\ls{\left[}
\def\rs{\right]}
\def\lc{\left\{}
\def\rc{\right\}}

\def\p{\partial}

\def\S{\Sigma}

\def\s{\sigma}

\def\O{\Omega}

\def\a{\alpha}
\def\b{\beta}
\def\g{\gamma}

\def\ad{{\dot \alpha}}
\def\bd{{\dot \beta}}
\def\gd{{\dot \gamma}}
\newcommand{\ft}[2]{{\textstyle\frac{#1}{#2}}}
\def\ib{{\overline \imath}}
\def\jb{{\overline \jmath}}
\def\Re{\mathop{\rm Re}\nolimits}
\def\Im{\mathop{\rm Im}\nolimits}
\def\trace{\mathop{\rm Tr}\nolimits}
\def\rmi{{ i}}

\def\N{\mathcal{N}}

\newcommand{\SU}{\mathop{\rm SU}}
\newcommand{\SO}{\mathop{\rm SO}}
\newcommand{\U}{\mathop{\rm {}U}}
\newcommand{\USp}{\mathop{\rm {}USp}}
\newcommand{\OSp}{\mathop{\rm {}OSp}}
\newcommand{\Symp}{\mathop{\rm {}Sp}}
\newcommand{\Sl}{\mathop{\rm {}S}\ell }
\newcommand{\Gl}{\mathop{\rm {}G}\ell }
\newcommand{\Spin}{\mathop{\rm {}Spin}}
\newcommand*\mybluebox[1]{\colorbox{blue!20}{\hspace{1em}#1\hspace{1em}}}

\newlength{\Lnote}
\newcommand{\notte}[1]
     {\addtolength{\leftmargini}{4em}
    \settowidth{\Lnote}{\textbf{Note:~}}
    \begin{quote}
    \rule{\dimexpr\textwidth-2\leftmargini}{1pt}\\
    \mbox{}\hspace{-\Lnote}\textbf{Note:~}%
    #1\\[-0.5ex] 
    \rule{\dimexpr\textwidth-2\leftmargini}{1pt}
    \end{quote}
    \addtolength{\leftmargini}{-4em}}
\def\hc{c.c.}

\numberwithin{equation}{section}

\allowdisplaybreaks

\allowbreak



\thispagestyle{empty}
\begin{flushright}

\end{flushright}
\vspace{35pt}


\begin{center}
	    {  \bf{\Large{  Integer dual dimensions in scale-separated AdS$_3$ 
	    
        \vspace{0.25cm} 
        
        from massive IIA  
	    }}}

		\vspace{50pt}

    	{\Large Fotis~Farakos{}$^a$~and~George~Tringas{}$^b$}

		\vspace{30pt}

{
\large 
{}$^a$
Physics Division, National Technical University of Athens,\\ 15780 Zografou Campus, Athens, Greece\\
\vspace{2mm}
{}$^b$ 
Department of Physics, Lehigh University,\\ 16 Memorial Drive East, Bethlehem, PA 18018,
USA
}


\vspace{1.3cm}

E-mails: fotis.farakos@gmail.com, georgios.tringas@lehigh.edu

\vspace{1.3cm}

{ABSTRACT}

\end{center}

We study supersymmetric scale-separated AdS$_3$ flux vacua of massive IIA on G2 orbifolds with smeared orientifold planes. 
We consider two types of $T^7/Z_2^3$ orbifolds which, with appropriate flux choices, 
yield integer dual dimensions for the operators corresponding to the closed string scalar fields in the dual CFT. 
As with all other known examples, the dual conformal dimensions are only parametrically close to integer values.

\thispagestyle{empty} 

\setcounter{page}{0}

\newpage


{\hypersetup{hidelinks}
\tableofcontents
}

\baselineskip 5.8 mm 

\setcounter{footnote}{0}


\section{Introduction}

Setting aside the phenomenological relevance of parametrically scale-separated solutions in string theory, one can investigate this question in its own right.
Such a theoretical question necessitates an investigation into the existence and self-consistency of these solutions across various dimensions with varying degrees of underlying supersymmetry.
Even for supersymmetric AdS vacua, a complete classification remains elusive, and the self-consistency of existing constructions is still under debate.
For contemporary reviews and an extended discussion see \cite{Lust:2019zwm,VanRiet:2023pnx,Coudarchet:2023mfs}.

Here, our focus is on classical supersymmetric vacua of string theory exhibiting parametric scale separation. 
As things currently stand, the only known 4D vacua with parametric scale separation are found either in massive IIA supergravity with smeared orientifold planes \cite{DeWolfe:2005uu,Camara:2005dc} (see \cite{Behrndt:2004km,Behrndt:2004mj,Derendinger:2004jn, Villadoro:2005cu} for earlier related constructions that share some constituent ingredients and features; and \cite{Marchesano:2019hfb} for recent extensions) or in massless IIA/M-theory with torsion in the internal space \cite{Cribiori:2021djm,VanHemelryck:2024bas}; both constructions preserve N=1 supersymmetry.
The constructions within massive IIA have long been under debate (see, e.g., \cite{McOrist:2012yc,Gautason:2015tig,Font:2019uva,Montero:2024qtz}), but their consistency remains to be definitively determined.
For the moment, the backreaction of localized orientifolds has been studied to first order \cite{Junghans:2020acz,Marchesano:2020qvg,Andriot:2023fss} and second order \cite{Emelin:2024vug}, yielding encouraging results.
Furthermore, there are specific instances of well-understood solutions where the backreaction can be consistent with the smeared limit \cite{Blaback:2010sj,Baines:2020dmu}, but this property appears to be compromised when the orientifolds are intersecting \cite{Bardzell:2024anh}.
Crucially, it is now understood that massive IIA constructions do not require intersecting orientifolds when the orbifold singularities are blown up \cite{Junghans:2023yue}.

As far as extensions go, 4D solutions with anisotropic internal spaces have also been studied in \cite{Carrasco:2023hta,Tringas:2023vzn}, while a possible reconciliation of scale-separated solutions related to \cite{DeWolfe:2005uu} with the distance conjecture has been explored, for example, in \cite{Shiu:2022oti,Shiu:2023bay,Palti:2024voy} and \cite{Tringas:2023vzn}.
However, solutions with higher supersymmetry appear to be excluded based on general supergravity swampland arguments \cite{Cribiori:2022trc,Cribiori:2023gcy}, although a direct proof from flux compactifications would also be valuable. 
In the meantime, it has been observed that the putative dual CFT of such IIA vacua includes operators with integer dimensions \cite{Aharony:2008wz,Conlon:2021cjk,Apers:2022zjx,Apers:2022tfm,Quirant:2022fpn,Ning:2022zqx,Apers:2022vfp}.
Why this property should hold remains an open question, and it is known that it will inevitably be disrupted once higher-order corrections are taken into account \cite{Plauschinn:2022ztd}. 
In other words, the notion of having integer conformal dimensions for the dual CFT operators makes sense only in the asymptotic formal limit of infinite volume and $g_s\to0$ such that the corrections are eliminated. 
What we will observe here is a possible interplay between the consistency of these scale-separated solutions and the existence of integer dual dimensions (always in the aforementioned asymptotic sense), even though one does not necessarily imply the other.

Moving beyond 4D solutions, one can explore either higher or lower dimensions.
Supersymmetric systems in more than four dimensions inherently require greater supersymmetry, which in turn obstructs scale separation \cite{Cribiori:2023ihv}.
On the other hand, in three or fewer dimensions, some possibilities remain. As for 2D cases, the only viable option at present appears to be minimal supersymmetry \cite{Cribiori:2024jwq} (see also attempts in \cite{Lust:2020npd}), though a concrete realization of such a minimal construction has yet to be found. 
In three dimensions with minimal supersymmetry, the setup is richer and more promising. Flux vacua of massive IIA on G2-orbifolds have been constructed in \cite{Farakos:2020phe}, utilizing smeared sources to achieve scale separation, while the backreaction of localized sources has been examined to first order \cite{Emelin:2022cac}.
3D vacua of IIB with scale separation have been identified in \cite{Arboleya:2024vnp,VanHemelryck:2025qok} (non-supersymmetric and supersymmetric, respectively), which, interestingly, also feature integer dual dimensions. 
Additional properties of 3D type II flux vacua, with an emphasis on the search for scale separation, have been studied in \cite{Emelin:2021gzx, VanHemelryck:2022ynr, Farakos:2023nms, Farakos:2023wps}.

Our specific focus here will be on the G2-orientifold construction of \cite{Farakos:2020phe}, which allows for a highly general choice of fluxes. The particular flux choice made in \cite{Farakos:2020phe} did not yield integer dual dimensions. Moreover, the blow-up of the orbifold singularities was not further investigated, and it is significant that the orbifold used in \cite{Farakos:2020phe} was not of the Joyce type \cite{Joyce:2002eb}. Here, we find that when studying orbifolds of the Joyce type, integer dual dimensions naturally emerge (always in the asymptotic limit that eliminates any corrections). These orbifolds are special in that they possess a known de-singularization, which Joyce used to construct the first compact G2-spaces \cite{Joyce:2002eb}. This makes such constructions even more intriguing, as it raises the possibility that, once de-singularized, the orientifolds may no longer be intersecting. Of course, as we will show, the orbifold employed in \cite{Farakos:2020phe} can also yield integer dual dimensions for specific flux choices; however, these fluxes precisely match those dictated by the Joyce orbifold.

\section{G2 orientifolds}

Compactifications on spaces with exceptional holonomy are well-studied in string theory because they lead to small amounts of preserved supersymmetry. 
Spaces with G2 holonomy preserve one-eighth of the original supersymmetry, leading to a vacuum with four Killing spinors in the case of massive IIA supergravity. If in addition we include orientifold planes supersymmetry can be further reduced. 
Here, we will introduce O2- and O6-planes in a way that preserves only two Killing spinors, corresponding to minimal supersymmetry in 3D.

One way a G2 space can be constructed is by starting with a toroidal orbifold 
\be
X_7 = \frac{T^7}{Z_2 \times Z_2 \times Z_2} \,, 
\ee 
and then de-singularizing it. 
Once that is done one has a smooth G2 manifold. 
Here, we will not treat the blowing-up of the singularities; instead, we will stay focused on the untwisted sector. 
However, 
we will distinguish the manner in which the singularities are to be treated; 
this will distinguish the type of orbifold that we will call ``singular'', from the ``Joyce orbifold''. 
The general orbifold we will work with is generated by the following $Z_2$ involutions 
\be
\begin{aligned}
\label{Z2s}
\Theta_\alpha : y^i & \to (-y^1, -y^2, -y^3, -y^4, y^5, y^6, y^7) \,, 
\\[0.5mm]
\Theta_\beta : y^i & \to (-y^1, c-y^2, y^3, y^4, -y^5, -y^6, y^7) \,,
\\[0.5mm]
\Theta_\gamma : y^i & \to (c-y^1, y^2, c-y^3, y^4, -y^5, y^6, -y^7) \,, 
\end{aligned}
\ee
where $c$ is either vanishing or non-vanishing half-integer. 
Note that the toroidal coordinates already satisfy the periodic identification $y^i \sim y^i + 1$. 
More specifically, the two options for $c$ are:
\begin{itemize}

\item $c=0$ is what we will refer to as the singular orbifold case which is used in \cite{Farakos:2020phe}. 

\item $c=1/2$ is what we will refer to as Joyce orbifold \cite{Joyce:2002eb}. 

\end{itemize}
Both of these choices lead to integer conformal dimensions for appropriate flux choices as we will see later on. 
To proceed we notice that the combinations of the above involutions give 
\be
\begin{aligned}
\Theta_{\alpha}\Theta_{\beta} : y^i & \to (y^1, c+ y^2, -y^3, -y^4, -y^5, -y^6, y^7) \,, 
\\[0.5mm]
\Theta_{\beta}\Theta_{\gamma} : y^i & \to (c+y^1, c-y^2, c-y^3, y^4, y^5, -y^6, -y^7) \,,
\\[0.5mm]
\Theta_{\gamma}\Theta_{\alpha} : y^i & \to (c+y^1, -y^2, c+y^3, -y^4, -y^5, y^6, -y^7) \,, 
\\[0.5mm]
\Theta_{\alpha}\Theta_{\beta}\Theta_{\gamma} : y^i & \to (c-y^1, c+y^2, c+y^3, -y^4, y^5, -y^6, -y^7) \,. 
\end{aligned}
\ee 
We observe that when $c = 0$, the fixed points of each of the seven involutions form 16 copies of $T^3$, which are further identified among themselves due to the action of the other involutions.
In contrast, when $c=1/2$, the fixed points are 16 $T^3$ only for the $\Theta_{\alpha}$, $\Theta_{\beta}$ and $\Theta_{\gamma}$, which can be blown-up to give a smooth G2. 
The reason that for $c=1/2$ we have less singularities is because the $\Theta_{\alpha}\Theta_{\beta}$, 
$\Theta_{\alpha}\Theta_{\gamma}$, 
$\Theta_{\beta}\Theta_{\gamma}$ and 
$\Theta_{\alpha} \Theta_{\beta}\Theta_{\gamma}$ are all freely acting. 
Let us denote the full set of these involutions as $\Gamma=\{\Theta_{\alpha},\Theta_{\beta},\Theta_{\gamma}\}$, 
so that the orbifold is given by $T^7/\Gamma$.

The next step is to include orientifold planes. 
To deduce a self-consistent set of orientifold planes we start by introducing an O2 with $Z_2$ involution 
\be
\label{O2}
\sigma: y^i \to - y^i \,. 
\ee
This in turn generates a web of mirror orientifolds which are all O6-planes and are located at the fixed points of the involutions 
$\sigma \Gamma$. 
For completeness we write down in any case the full set of the involutions which act as 
\be
\begin{aligned}
\Theta_\alpha\sigma : y^i & \to (y^1, y^2, y^3, y^4, -y^5, -y^6, -y^7) \,, 
\\[0.5mm]
\Theta_\beta\sigma : y^i & \to (y^1, c+y^2, -y^3, -y^4, y^5, y^6, -y^7) \,,
\\[0.5mm]
\Theta_\gamma\sigma : y^i & \to (c+y^1, -y^2, c+y^3, -y^4, y^5, -y^6, y^7) \,, 
\\[0.5mm] 
\Theta_{\alpha}\Theta_{\beta}\sigma : y^i & \to (-y^1, c -y^2, y^3, y^4, y^5, y^6, -y^7) \,, 
\\[0.5mm]
\Theta_{\beta}\Theta_{\gamma}\sigma : y^i & \to (c-y^1, c+y^2, c+y^3, -y^4, -y^5, y^6, y^7) \,,
\\[0.5mm]
\Theta_{\gamma}\Theta_{\alpha}\sigma : y^i & \to (c-y^1, y^2, c-y^3, y^4, y^5, -y^6, y^7) \,, 
\\[0.5mm]
\Theta_{\alpha}\Theta_{\beta}\Theta_{\gamma}\sigma : y^i & \to (c+y^1, c-y^2, c-y^3, y^4, -y^5, y^6, y^7) \,. 
\end{aligned}
\ee
We see that depending on the value of $c$ the type of O6 planes changes because some involutions are freely acting, 
therefore we will discuss these two cases distinctly.

\subsection{Joyce G2 orbifold} 

Here $c=1/2$ and therefore we can see that only three of the seven involutions have fixed points which are then interpreted as O6-planes. 
The non-freely acting involutions are 
\be
\begin{aligned}
\label{O6-1/2}
\Theta_\alpha\sigma : y^i & \to (y^1, y^2, y^3, y^4, -y^5, -y^6, -y^7) \,, 
\\[0.5mm]
\Theta_{\alpha}\Theta_{\beta}\sigma : y^i & \to (-y^1, 1/2 -y^2, y^3, y^4, y^5, y^6, -y^7) \,, 
\\[0.5mm]
\Theta_{\gamma}\Theta_{\alpha}\sigma : y^i & \to (1/2-y^1, y^2, 1/2-y^3, y^4, y^5, -y^6, y^7) \,. 
\end{aligned}
\ee
These involutions then generate a web of O6-planes which together with the O2 are located at the positions 
\begin{align}
\begin{pmatrix} 
&{\rm O}2:\quad & - & - & - & - & - & - & -  \\
&{\rm O}6_{\alpha}:\quad & \times & \times & \times & \times & - & - & -  \\
&{\rm O}6_{\alpha\beta}:\quad & - & \sim & \times & \times & \times & \times & -  \\ 
&{\rm O}6_{\gamma\alpha} :\quad & \sim & \times & \sim & \times & \times & - & \times  
\end{pmatrix} \, . 
\end{align}
Here, we use $\times$ to indicate the internal directions that are parallel to the orientifold planes, 
i.e. directions belonging to their world-volume. 
With the symbol $-$ we indicate directions where the orientifold is localized at the positions $0,1/2$, 
while with $\sim$ we refer to directions where the orientifold is localized at the positions $1/4,3/4$. 
Unavoidably, however, the orientifolds will be smeared along the $-$ and $\sim$ directions. 
The O2-planes and the D2-branes are also smeared as well.

Specifically, for the massive IIA setup, we will closely follow the construction of \cite{Farakos:2020phe}. In that framework the only fluxes involved are $F_4$ and $H_3$. For complete details on the full setup, we refer the reader to \cite{Farakos:2020phe}.
The tadpole conditions that have to be satisfied arise from the integral of the following non-trivial Bianchi identities
\be
0 = H_3 \wedge F_0 + j_{{\rm O}6} + j_{{\rm D}6} \,, 
\ee
and 
\be
0 = H_3 \wedge F_4 + j_{{\rm O}2} + j_{{\rm D}2}   \,, 
\ee
where $j_{{\rm O}6}$ refers to the (smeared) source current corresponding to the O6-plane in the internal space and similarly for the other localized objects. The charge and number of the local objects are encoded in the currents.
At this point, we choose to cancel the O2-plane with the appropriate number of D2-branes, imposing $j_{{\rm O}2} + j_{{\rm D}2} = 0$. Additionally, since we do not introduce D6-branes, the fluxes should be arranged such that
\be\label{tad3}
H_3 \wedge F_4 =0 \quad, \quad H_3 \wedge F_0 = - j_{{\rm O}6}  \,. 
\ee
Since we are working on a G2 orbifold we can use the basis of harmonic 3-forms 
\be
\label{3basis}
\Phi_i = \left(- dy^{567},  dy^{127}, dy^{136}, - dy^{347}, - dy^{235}, dy^{145}, dy^{246} \right) \ , \quad i = 1 , \dots, 7 \, , 
\ee
which also have the dual basis for harmonic 4-forms 
\be
\label{4basis}
\Psi_i = \left(- dy^{1234},  dy^{3456}, dy^{2457}, -dy^{1256}, - dy^{1467}, dy^{2367}, dy^{1357} \right) \ , \quad i = 1 , \dots, 7 \, . 
\ee
Note that these forms satisfy the orthogonality condition $\int \Phi_i \wedge \Psi_j = \delta_{ij}$. 
With this notation we see that the three non-trivial $j_{{\rm O}6}$ currents are oriented along the $\Phi_1$, 
$\Phi_2$, and $\Phi_3$ directions. 
This means our flux choices are unavoidably 
\be\label{H3}
H_3 = h (\Phi_1 + \Phi_2 + \Phi_3)\quad,\quad F_0 = m \,, 
\ee
while for the $f$ we have more freedom but we will initially consider the simplest option which is 
\be\label{F4}
F_4 = - f (\Psi_4 + \Psi_5 + \Psi_6 + \Psi_7) \,. 
\ee
We can also ensure that, due to this flux ansatz on the specific cycles, the only non-trivial flux equation of motion that arises from the NSNS 2-form, $H_3 \wedge \star_7 F_4 = 0$, is satisfied because the wedge product of these basis elements identically vanishes. 
Thus, using the flux ansatz and the proper arrangement of fluxes and sources, we can satisfy the tadpoles, trivially solve the flux equations, and consequently keep the $F_4$ flux unbounded.

For completeness let us note that once flux quantization and orientifold charge is taken into account we have 
\be
h =  (2\pi)^2 K \quad  , \quad m=(2\pi)^{-1} M \quad ,\quad KM = 16 \quad , \quad f = (2\pi)^3 N \,, 
\ee 
where $N,K,M\in\mathbb{Z}$ and $\alpha'=1$.

We observe that the orientifold planes in this setup are intersecting. 
This implies that evaluating the backreaction of the localized sources on the smeared solution requires higher-order corrections to ensure self-consistency. 
However, since the orbifold singularities can be resolved, one should first check whether the sources remain intersecting after the blow-ups. 
If the sources cease to be intersecting after the resolution, similar to what happens in Calabi--Yau orientifolds as shown in \cite{Junghans:2023yue}, then this setup certainly calls for further investigation. 

\subsection{Singular G2 orbifold}

When $c=0$, all $\sigma \Gamma$ involutions have fixed points. 
This produces a web of mutually supersymmetric intersecting orientifold planes located as follows 
\begin{align}
\begin{pmatrix} 
&{\rm O}6_{\alpha}:\quad & \times & \times & \times & \times & - & - & -  \\
&{\rm O}6_{\beta} :\quad &\times & \times & - & - & \times & \times & -  \\
&{\rm O}6_{\gamma}:\quad &\times & - & \times & - & \times & -& \times  \\
&{\rm O}6_{\alpha\beta}:\quad & - & - & \times & \times & \times & \times & -  \\ 
&{\rm O}6_{\beta\gamma} :\quad & - & \times & \times & - & - & \times & \times  \\ 
&{\rm O}6_{\gamma\alpha} :\quad &- & \times & - & \times & \times & - & \times  \\ 
&{\rm O}6_{\alpha\beta\gamma}:\quad &\times & - & - & \times & - & \times & \times 
\end{pmatrix} \, . 
\end{align}
To satisfy the tadpole conditions for this setup, we can use either fluxes or D6-branes.
The choice of D6-branes is made to ensure the same flux configuration as in the Joyce-type orbifold setup, namely
\begin{align}
\begin{pmatrix} 
&{\rm D}6_{\beta} :\quad &\times & \times & - & - & \times & \times & -  \\
&{\rm D}6_{\gamma}:\quad &\times & - & \times & - & \times & -& \times  \\
&{\rm D}6_{\gamma\alpha} :\quad &- & \times & - & \times & \times & - & \times  \\ 
&{\rm D}6_{\alpha\beta\gamma}:\quad &\times & - & - & \times & - & \times & \times 
\end{pmatrix} \, . 
\end{align}
This setup produces the same untwisted closed string sector as the one with the Joyce orbifold.
However, given the presence of D6-branes and a non-trivial $H_3$ background, one might be concerned about potential Freed--Witten anomalies.
Here, we find that the pullback of $H_3$ to the D6-branes vanishes, ensuring that no such issue arises.

\section{Scale-separated AdS$_3$ from massive type IIA}   

We have seen that there are two different G2-orbifolds that are relevant for us, 
both admitting the same flux choices. 
This happens due to the identical net smeared charge that should be canceled in the Bianchi identities, 
despite differences in the smeared sources (O6-planes and D6-branes). 
We will now study these two cases together, as they lead to the same behavior in the universal untwisted closed string sector. 
In doing so, we will observe that they result in integer dual dimensions.

In this section, however, we will see that there is a flat direction that remains undetermined in these AdS background, 
which is a threat to the solution once higher order corrections unavoidably are taken into account. 
In the next section we will take care of this flat direction and provide full moduli stabilization. 
The reason we first work with this example is because integer conformal dimensions show up readily at the classical level. 
Of course, unavoidably, as in other solutions higher order corrections spoil such property even if full moduli stabilization is achieved \cite{Plauschinn:2022ztd}.

\subsection{The 3D effective theory}

To examine the properties of the effective theory arising from the compactification of massive type IIA on G2-orientifolds, 
we begin with the following 10D metric ansatz in the Einstein frame 
\begin{equation}
    ds^2_{10} = \frac{(2 \pi)^{14}}4 e^{2\alpha\upsilon} ds^2_3+ e^{2\beta\upsilon} d\tilde{s}^2_7 \,, 
\end{equation}    
where $\upsilon$ is a real scalar that characterizes the size of the internal space volume in the Einstein frame.
The internal space volume is given by  
\begin{equation}  
\text{vol}(X_7) = e^{7 \beta \upsilon} \text{vol}(\tilde{X}_7) \,,  
\end{equation}  
where the tilde denotes a unit-volume internal space and $\text{vol}(\tilde{X}_7)=1$ its corresponding volume. 
The coefficients $\alpha^2 = 7/16$ and $\beta = -\alpha/7$ are chosen to ensure canonical kinetic terms for the universal moduli, the volume $\upsilon$ and the dilaton $\phi$, after dimensional reduction, and for convenience, we perform the following orthonormal redefinitions of these universal scalar fields 
\begin{equation}
    \frac{x}{\sqrt{7}}=-\frac{3}{8}\phi+\frac{\beta}{2}\upsilon\quad ,\quad y=-\frac{21}{2}\beta\upsilon-\frac{1}{4}\phi \,.
\end{equation}
The superpotential of the 3D N=1 effective theory, arising from general G2-compactification with the field and source content described in the previous section, was found in \cite{Farakos:2020phe} to have the following form for general flux choices 
\begin{equation}\label{superpotential}
    \frac{P}{(2\pi)^7}=
    \frac{m}{2}e^{\frac{1}{2}y-\frac{\sqrt{7}}{2}x}
    +\frac{1}{8}e^{y+\frac{x}{\sqrt{7}}}
    \int_7\star\Phi\wedge H_3\,\text{vol}(X_7)^{-\frac{4}{7}}
    +\frac{1}{8}e^{y-\frac{x}{\sqrt{7}}}
    \int_7\Phi\wedge F_4\,\text{vol}(X_7)^{-\frac{3}{7}} \,.
\end{equation}
The fluctuations of the metric are given by $s^i = e^{3\beta\upsilon} \tilde{s}^i$, 
where $\tilde s^i$ are fluctuations for the unit-volume space (i.e. shape moduli) with $i=1\,\dots,7$, 
and the associative 3-form of the G2 space is $\Phi=s^i\Phi_i$. 
From now on, due to the unit-volume choice, we express one of the internal space fluctuations in terms of the rest, with
\begin{equation}\label{unit}
    \text{vol}(\tilde{X}_7)=1\,\rightarrow\,\tilde{s}^7=1/\prod_{a=1}^6\tilde{s}^a \quad,\quad a=1,\,\dots,6\,. 
\end{equation} 
The scalar-field lagrangian of the 3D effective theory has the following form
\begin{equation}\label{lagrangian}
    e^{-1}\mathcal{L}
    =
    \frac{1}{2}R_3
    -\frac{1}{4}\left(\partial x\right)^2
    -\frac{1}{4}\left(\partial y\right)^2
    -G_{ij} \partial\tilde{s}^i\partial\tilde{s}^j
    -V(\tilde{s}^{i},x,y) \,,
\end{equation}
where the shape-moduli kinetic terms and the total scalar potential are given by
\begin{equation} 
G_{ij} 
= \frac{1}{4}\text{vol}(\tilde{X})^{-1}\int_7\Phi_i\wedge \tilde{\star} \Phi_j 
= \frac{1}{4 (\tilde s^i)^2} \delta_{ij} 
\ , \  
    V(\tilde{s}^{i},x,y) = G^{ij}P_jP_j + 4 (P_x^2 + P_y^2 - P^2)  
    \,.
\end{equation}
Once that replacement of the $\tilde s^7$ is made the kinetic terms of the remaining eight real scalars take the form 
\be
e^{-1}\mathcal{L}_{kin.} = - \frac12 K_{AB} \partial \varphi^A \partial \varphi^B \qquad A,B = 1,\dots, 8 \, , 
\ee
with $\varphi^A=x,y,\tilde s^1,\dots, \tilde s^6$, and 
\be
K_{AB} 
= 
\begin{pmatrix} 
1/2 & 0 & 0  \\
0 & 1/2 & 0   \\
0 & 0 & k_{ab} 
\end{pmatrix} 
\quad , \quad 
k_{ab} = \frac{1 + \delta_{ab}}{2 \tilde s^a \tilde s^b} 
\quad , \quad  a,b=1,\dots,6 
\,. 
\ee

\subsection{Supersymmetric background and dual dimensions}

We can now analyze the supersymmetric vacua of the 3D N=1 theory by first evaluating the integrals of the superpotential \eqref{superpotential}, which takes the form
\be
\label{Pint}
\frac{P}{(2 \pi)^7 } \! 
= 
\frac{m}{8}e^{\frac{y}{2}-\frac{\sqrt{7}}{2}x} 
+\frac{h}{8} e^{y + \frac{x}{\sqrt{7}}} \left[ \frac{1}{\tilde s^1} + \frac{1}{\tilde s^2} +\frac{1}{\tilde s^3} \right] 
-\frac{f}{8}e^{y-\frac{x}{\sqrt{7}}}\left(\tilde{s}^4+\tilde{s}^5+\tilde{s}^6+\frac{1}{\prod_{a=1}^6\tilde{s}^a} \right) \! , 
\ee
with $f>0$, $m>0$ and $h>0$. 
Then, the extremization reduces to the eight equations, $P_x = 0 =  P_y = P_{\tilde{s}^a}=0$, which are directly solved by setting 
\be
\label{tsxy}
\tilde s^a = 1 \quad , \quad 
x = \frac{\sqrt{7}}{2} \log \left[ \frac{f}{h} \right] \quad , \quad 
y = 2 \log \left[ \frac{h^{5/4} m}{2 f^{9/4}} \right] \,. 
\ee
We then find the string coupling in terms of the fluxes 
\be
g_s = e^\phi = \exp \left[- \frac{3 \sqrt 7}{8} x - \frac18 y \right] = \frac{2^{1/4} h}{f^{3/4} m^{1/4} }  \,, 
\ee
which can be small for large $f$. 
Since higher order corrections come in the string frame, 
we evaluate the string-frame volume to ensure that we remain in the large volume regime, which is given by 
\be
\text{vol}_S(X_7) = g_s^{7/4} \text{vol}(X_7) = \left( \frac{2f}{m} \right)^{7/4} \,. 
\ee
From this, we can verify that taking $f$ large leads to weak coupling and large volume. Checking the tadpole cancellation condition, we see that there is no restriction on the value of $f$, allowing us to take it parametrically large.

We can now examine scale separation. First, we verify that our vacuum is indeed AdS by evaluating the vacuum energy on the solution
\be
V = - 4 P^2 = - \frac{64 \pi^{14} h^6 m^4}{f^8} \,.  
\ee
From \cite{Farakos:2020phe}, we know that to determine whether scale separation is realized we need to evaluate
\be
\frac{L_{KK}^2}{L_{AdS}^2} = e^{16 \beta v} |V| = \frac{512 \pi^{14} h^2 m }{f} \,. 
\ee
Thus, we verify that in the weak coupling and large volume limit, parametric scale separation is realized. 
Indeed, when $f \to \infty$ we see that $L_{KK}^2 / L_{AdS}^2  \to 0$.

Now we turn to the evaluation of the masses and then to the conformal dimensions in the dual CFT. 
To find the masses of the moduli in units of the cosmological constant, 
we calculate the eigenvalues of the normalized Hessian which is given by the following matrix 
\begin{equation}
\label{eigen}
m_A^2 L^2 
= \text{Eigen}\left[\langle K_{AB}\rangle^{-1}\frac{\langle V_{BC}\rangle}{\vert\langle V\rangle\vert}\right] 
= \left\{48, 8, 8, 8, 8, 0, 0, 0\right\} 
    \quad , \quad A=1,\dots,8  \,, 
\end{equation}
where the indices $B,C$ in $V_{BC}$ correspond to derivatives with respect to the $\varphi^A$ scalars. 
Note that one of the zero modes of the Hessian is a flat direction, which we will identify and stabilize in the next section. 
Moving on, we can evaluate the dimensions of the dual operators in the putative dual CFT 
\begin{equation}
\label{dual-op}
\Delta_A = 1 + \sqrt{ 1 + m_A^2 L^2 } =  \left\{8,4,4,4,4,2,2,2\right\} \,. 
\end{equation}
We conclude that the conformal dimensions of the field theory operators dual to the scalars take integer values.

\subsection{An RR-flux choice with anisotropy} 

The setup we are studying here certainly allows for a lot of freedom, 
and it is neither possible here nor our aim to cover all the options. 
However, it is worth to investigate what happens in other examples of flux choices, 
which give rise to anisotropic solutions. 
To this end we will take the $F_4$ flux to have the form 
\be
f = - f (\Psi_4 + \Psi_5 + \Psi_6 + n \Psi_7) \quad , \quad f,n > 0 \,. 
\ee
Here, $n$ needs only to be a rational number and as long as $n \times N$ is an integer we will maintain flux quantization 
(we remind the reader that $f = (2\pi)^3 N$). 
Notice also that there is nothing special with inserting $n$ in front of $\Psi_7$, 
we could have equally well inserted it in front of $\Psi_4$ or one of the other harmonic four-forms. 
This eventually just rotates the anisotropy in the internal space. 

In this setup the superpotential becomes 
\be
\frac{P}{(2 \pi)^7 } 
= \! 
\frac{m}{8}e^{\frac{y}{2}-\frac{\sqrt{7}}{2}x}
+\frac{h}{8} e^{y + \frac{x}{\sqrt{7}}} \left[ \frac{1}{\tilde s^1} + \frac{1}{\tilde s^2} +\frac{1}{\tilde s^3} \right] 
-\frac{f}{8}e^{y-\frac{x}{\sqrt{7}}}\left(\tilde{s}^4+\tilde{s}^5+\tilde{s}^6+\frac{n}{\prod_{a=1}^6\tilde{s}^a} \right) \!, 
\ee
still with $m>0$ and $h>0$. 
The solution to the extremization is now 
\be
\tilde s^a = n^{1/7} 
\quad , \quad 
x = \frac{\sqrt{7}}{2} \log \left[ \frac{f n^{2/7}}{h} \right] 
\quad , \quad 
y = 2 \log \left[ \frac{h^{5/4} m}{2 f^{9/4} n^{1/2}} \right] \,, 
\ee
and we have an anisotropic solution because 
\be
\tilde s^7 = \frac{1}{\prod_{a=1}^6\tilde{s}^a} = n^{-6/7} \,. 
\ee
The vacuum remains a supersymmetric AdS$_3$. 
For the string-frame volume and the string coupling we have 
\be
\text{vol}_S(X_7) = n^{1/4}  \left( \frac{2f}{m} \right)^{7/4} 
\quad , \quad 
g_s =  n^{-1/4}  \frac{2^{1/4} h}{f^{3/4} m^{1/4} } \,, 
\ee
while the condition for scale separation becomes 
\be
\frac{L_{KK}^2}{L_{AdS}^2} = \frac{512 \pi^{14} h^2 m }{f n^{3/7}} \,. 
\ee
This means we can have parametric scale separation at large volume and weak coupling when the unbounded RR-flux $f$ becomes parametrically large, under the assumption that $n$ takes some moderate values. 
Once we evaluate the dimension of the dual operators we see that they also have the values \eqref{dual-op}, 
independent of the value of $n$.
This property, wherein the dimensions of the dual operators remain unchanged in anisotropic cases that induce mass variations, was also noted in \cite{Farakos:2023nms}.

To examine the behavior of anisotropy, we examine the radii. 
We find that the Einstein-frame radii behave as 
\be
r^1 = r^3 = r^5 = r^7 = \left( \frac{8 f^7 n^5}{m^3 h^4} \right)^{1/16} 
\quad , \quad 
r^2 = r^4 = r^6 = \left( \frac{8 f^7 }{n^3 m^3 h^4} \right)^{1/16}  \,. 
\ee
From this, we observe that varying $n$ induces an anisotropic internal space.
However, to stay at large string-frame radii one should respect the conditions 
\be
\left( \frac{2 fn}{m} \right)^{1/4} \gg 1 
\quad , \quad 
\left( \frac{2 f}{m n} \right)^{1/4} \gg 1 \,, 
\ee
on top of the condition for weak string coupling. 
We see that for parametrically large $f$ there is quite some freedom for the choice of anisotropy.

\section{More flux and full moduli stabilization}

A notable property of the mass matrix \eqref{eigen} is that three of its eigenvalues vanish. In fact, one can show that one of the scalars remains a modulus of the supersymmetric background. 
Indeed, if we analyze the superpotential we can observe that it remains invariant under the following shift 
\be
\label{shifts}
\begin{aligned}
x & \to x + \frac{a}{\sqrt 7} \, , 
\\
y & \to y + a \, , 
\\
\tilde s^{1,2,3} & \to e^{\frac{8a}{7}} \times \tilde s^{1,2,3} \, , 
\\
\tilde s^{4,5,6} & \to e^{- \frac{6a}{7}} \times \tilde s^{4,5,6} \, . 
\end{aligned}
\ee
This implies that one of the scalars has an unspecified value, which will inevitably be influenced by higher-order terms or non-perturbative corrections. 
One might suspect that the modulus responsible for \eqref{shifts} is an artifact of the smearing approximation. 
Indeed, once the localized sources are allowed to backreact, it seems unlikely that they do not somehow influence the modulus. 
The concern is that, once localized sources and higher-order terms are introduced, their impact on the solution could potentially eliminate the critical point.
Of course, it goes without saying that the property of integer conformal dimensions for the dual operators will certainly be lost once any form of corrections is included. 
This is a known effect that also occurs in other examples \cite{Plauschinn:2022ztd}. In other words, the actual integer values in any known setup are ultimately an artifact of truncating higher-order corrections, and are integer at best only up to parametrically small corrections. Moduli stabilization, however, must always be maintained; otherwise, we lose control over the solution.

\subsection{Flux choices and analytic solution}

With the above in mind, we aim to investigate whether it is still possible to have a classical supersymmetric AdS vacuum that supports full moduli stabilization and has dual conformal dimensions that are arbitrarily close to integer values. 
One way to achieve this is by removing some of the D2-branes and allowing the fluxes to contribute to tadpole cancellation, 
but also slightly generalizing our flux profile. 
This means we have 
\be
\int F_4 \wedge H_3 = (2 \pi)^5 (16 - N_{D2}) \,, 
\ee
all while remaining within the smeared approximation. 
The $H_3$ flux is unaltered, as does the Romans mass $F_0$. 
To be manifestly consistent with flux quantization let us right away set 
\be
H_3 = (2 \pi)^2 (\Phi_1 + \Phi_2 + \Phi_3) \quad , \quad F_0 = 16 (2 \pi)^{-1} \,. 
\ee
For the $F_4$ flux we take a more general profile compared to the previous section 
\be
\begin{aligned}
F_4 = (2 \pi)^3 \Big{[} 
& - N \left( \Psi_4 + \Psi_5 + \Psi_6 + (1+n) \Psi_7 \right) 
\\
& + Q (\Psi_1 + \Psi_2 + \Psi_3) 
+ G (\Psi_1 + \Psi_2 -2 \Psi_3) 
\Big{]} \,. 
\end{aligned}
\ee
We can already see that, with this flux ansatz (for non-zero values of the fluxes of course), 
the shift symmetry \eqref{shifts}, which previously existed in the superpotential, is broken. 
From now on for clarity we will specify some of the fluxes.  
In particular we will solve the tadpole cancellation condition as follows 
\be
3 Q = 16 - N_{D2} \quad  \to \quad  Q = 5 \ , \ N_{D2} = 1 \,. 
\ee 
We will keep these flux choices from now on. 
The limit we are interested in is 
\be
\label{limit}
N \gg  G \gg 1 \gg n \,, 
\ee
such that the $F_4$ flux orientation that leads to integer conformal dimensions dominates and we will make sure that $nN$ is always an integer. 

Now we turn to moduli stabilization. 
We extremize the superpotential with respect to the eight moduli $\varphi^A=x,y,\tilde s^a$, 
and once we write down these eight equations we observe that there exists an exact solution with the moduli taking the values 
\be
\tilde s^{1,2} = \sigma 
\quad , \quad 
\tilde s^{3} = \frac{1+n}{\sigma^2 \tau^4} 
\quad , \quad 
\tilde s^{4,5,6} = \tau \,, 
\ee
and 
\be
\begin{aligned}
x = & \frac{\sqrt 7}{2} \log\left[ \frac{45 (1+n)^2 \pi \sigma^2}{(1+n- \sigma^3 \tau^4)^2}  \right] \,, 
\\ 
y = & 2 \log\left[ \frac{
\left( \frac{(1+n)^2 \sigma^2}{(1+n- \sigma^3 \tau^4)^2} \right)^{3/4} (1+n- \sigma^3 \tau^4)^6
}{
675 \sqrt{3}  \pi^{21/4}  (1+n)^5 \sigma^5 5^{1/4}  (2+2n+\sigma^3 \tau^4) 
}  \right] \,. 
\end{aligned}
\ee
To have the above solution the moduli should satisfy the conditions 
\be
\label{NG}
\begin{aligned}
N = & \frac{15 (1+n) \sigma (2+2n+\sigma^3 \tau^4) }{2 \tau (1+n- \sigma^3 \tau^4)^2}  \,, 
\\
G = & \frac{5 (1+n+2\sigma^3 \tau^4) }{2 (1+n- \sigma^3 \tau^4)} \,. 
\end{aligned}
\ee
These equations can be solved numerically after selecting integer values for the fluxes. Indeed, if we take $G = 10^2$, $N = 4 \times 10^3$, and $n = 5 \times 10^{-2}$, we find $\sigma \simeq 0.95$ and $\tau \simeq 1.03$, while $g_s \ll 1$ and we are at large volume in both the string and Einstein frames. 
Of course the fact that we have identified a specific solution does not mean that other solutions are not possible.

If we further refine our flux choices we can find an analytic solution. 
Indeed, upon inspection it turns out to be convenient to set
\be 
\label{NGnAn}
N = \frac{2}{5} G^2 \quad , \quad n = \frac{5}{G} \,, 
\ee 
and to be consistent with flux quantization, $G$ must be divisible by 5. Furthermore, to simplify some expressions in the solution, it turns out that one must assume $G > 5$, as we will do from now on.
Then we find that \eqref{NG} is solved as follows 
\be
\sigma = \frac{2^{3/7} G (2G-5)^{1/7}}{\big((5+G)(5+2G)\big)^{4/7}}
\quad , \quad 
\tau =  \frac{\big((5+G)^3 (2G-5) (5+2G)^3\big)^{1/7}}{2^{4/7} G} 
\,. 
\ee 
To derive these solutions, one first determines that the two moduli satisfy the condition $G = -\frac{5}{2} (\sigma^3 \tau^4 - 1)^{-1}$, which is then used to obtain the explicit form of the solution. Naturally, substituting these solutions into the derivatives of the superpotential directly confirms that they correspond to extrema. We conclude that we have a fully analytic solution controlled by the integer $G$. 
For completeness we also present how the $x$ and $y$ moduli depend on the $G$ flux 
\be
x = \frac{\sqrt 7}{2} \log \left[ \frac{4 \pi}{5} 2^{6/7} G^2  
\left( 
\frac{
(5+G)^6 (2G-5)^2
}{(5+2G)^8}
\right)^{1/7}
\right] \,, 
\ee
and 
\be
y = 2 \log \left[ \frac{25 (5+2G) 5^{1/4} }{32 (5+G)^{3/2} G^{7/2} \pi^{21/4} \sqrt{2G-5}} \right]  \,. 
\ee
Furthermore, when calculating the normalized Hessian over the vacuum energy, we find that the determinant is
\be
\det\left[ \frac{V_{AB}}{|V|} \right] = - \frac{21 \times 2^{15} \times 5^4}{2^{1/7}} \times \frac{
(15 + 2 G) (5 + 6 G) (2G-5)^{2/7}
}{
(5+G)^{8/7} (5+2G)^{36/7} 
} \ne 0 \,. 
\ee
From here, we see that the determinant is always nonzero (this result holds for $G > 5$). The limiting case of a vanishing determinant is reached only as $G \to \infty$, which corresponds to the limit that yields integer dual dimensions. We conclude that full moduli stabilization is achieved without any flat directions. 

One can also check that for large $G$, we obtain weak coupling, large string-frame volume, and parametric scale separation
\be
g_s \sim G^{-3/2} 
\quad, \quad 
\text{vol}_S(X_7) \sim G^{7/2} 
\quad, \quad 
e^{16 \beta v} |V| \sim G^{-2} \,. 
\ee
We also observe that as $G$ increases, $n \to 0$, $\sigma \to 1$, and $\tau \to 1$, while $N \sim G^2$, ensuring that the condition \eqref{limit} is satisfied.

\subsection{Approaching the integer dual dimensions}

From now on, we focus on large values of $G$, which, in the limit $G \to \infty$, should yield the solution with integer dual dimensions. We aim to study the behavior near large $G$. With these choices, we find that the moduli are fixed at 
\be
\label{ST}
\sigma = 1 - \frac{65}{14 G} + {\cal O}(1/G^2) \quad , \quad \tau = 1 + \frac{20}{7G} + {\cal O}(1/G^2) \, , 
\ee
and 
\be
x = \frac{\sqrt{7}}{2} \log \left[ \frac{4 \pi}{5} G^2 \right] + {\cal O}(1/G) \quad , \quad 
y = \frac32 \log \left[ \frac{125}{64 \pi^7 G^6} \right] + {\cal O}(1/G) \,. 
\ee

Working around the solution at large $G$, one finds that the full normalized Hessian takes the form 
\be
\frac{V_{AB}}{|V|} = \begin{pmatrix}
    \frac{149}{7} & \frac{13}{\sqrt 7} & -\frac{10}{\sqrt 7} & -\frac{10}{\sqrt 7} & -\frac{10}{\sqrt 7} & 0 & 0 & 0  \\
    \frac{13}{\sqrt 7} & 5 & -2 & -2 & -2 & 0 & 0 & 0 \\ 
     -\frac{10}{\sqrt 7} & -2 & 4 & 4 & 4 & 4 & 4 & 4  \\
    -\frac{10}{\sqrt 7} & -2 & 4 & 4 & 4 & 4 & 4 & 4 \\ 
     -\frac{10}{\sqrt 7} & -2 & 4 & 4 & 4 & 4 & 4 & 4  \\
    0 & 0 & 4 & 4 & 4 & 8 & 4 & 4 \\ 
     0 & 0 & 4 & 4 & 4 & 4 & 8 & 4  \\
    0 & 0 & 4 & 4 & 4 & 4 & 4 & 8 
  \end{pmatrix} 
  + {\cal O}(1/G) \,, 
\ee
where the $\mathcal{O}(1/G)$ matrix can be explicitly computed numerically, though we do not require its details here, as we are only interested in the limit $G \to \infty$.  
For the kinetic matrix, we also have
\be
K_{AB} = \begin{pmatrix}
    \frac12 & 0 & 0 & 0 & 0 & 0 & 0 & 0  \\
    0 & \frac12 & 0 & 0 & 0 & 0 & 0 & 0  \\ 
     0 & 0 & 1 & \frac12 & \frac12 & \frac12 & \frac12 & \frac12  \\
   0 & 0 & \frac12 & 1 & \frac12 & \frac12 & \frac12 & \frac12 \\ 
    0 & 0 & \frac12 & \frac12 & 1 & \frac12 & \frac12 & \frac12  \\
    0 & 0 & \frac12 & \frac12 & \frac12 & 1 & \frac12 & \frac12 \\ 
    0 & 0 & \frac12 & \frac12 & \frac12 & \frac12 & 1 & \frac12  \\
    0 & 0 & \frac12 & \frac12 & \frac12 & \frac12 & \frac12 & 1 
  \end{pmatrix} 
  + {\cal O}(1/G) \,, 
\ee
where the $\mathcal{O}(1/G)$ matrix can be calculated explicitly or numerically. The upshot is that, in the formal limit $G \to \infty$, both the kinetic and the mass matrix asymptote to the values they have in our example of the previous section with dual integer dimensions. 
This means we have
\be
\Delta_A = 1 + \sqrt{ 1 + m_A^2 L^2 } \big{|}_{G\to \infty} =  \left\{8,4,4,4,4,2,2,2\right\}  \,, 
\ee
which means one can be parametrically close to integer conformal dimensions. 

The result we find can be easily explained by considering that 
\be
\frac{\langle P_{AB} \rangle}{\langle P\rangle} \Big{|}_{G\to \infty} 
\ \equiv \ 
\frac{\langle P^{int.}_{AB} \rangle}{\langle P^{int.} \rangle}  \,, 
\ee
where $P^{int.}$ is the superpotential \eqref{Pint} from the previous section. This occurs because $P$ differs from $P^{int.}$ only by $F_4$ flux terms, which, however, are chosen in such a way that they become subdominant at large $G$.
Meanwhile, we also have
\be
\langle \varphi_A \rangle |_{G\to \infty} \equiv \langle \varphi_A^{int.} \rangle |_{G\to \infty} \,. 
\ee
This means all moduli converge to the values that provide integer dual dimensions.

We conclude that the solution asymptotes to the solution with integer conformal dimensions of the dual operators, while $g_s \to 0$ and the volume diverges. This is exactly the same situation as in the other known solutions, since corrections to the masses inevitably exist. In other words, it is as if these models already enforce the existence of corrections at the classical level; corrections that are eliminated in the formal limit $g_s \to 0$, which corresponds to the limit $G \to \infty$.

\section{Summary and outlook}

In this work we have studied supersymmetric scale-separated AdS$_3$ flux vacua of massive IIA supergravity and identified the appearance of integer conformal dimensions. 
For the internal geometry, we provided two types of G2 toroidal orbifolds of the form $T^7/Z_2^3$. 
In particular, one of these orbifolds is exactly of the type used by Joyce to construct smooth compact G2-spaces, meaning it has a known de-singularization.
Inevitably, our analysis also involves intersecting orientifold planes which are smeared.
The backreaction of the localized profiles of these objects remains an open question and certainly deserves a careful study. 
However, it is conceivable that once the orbifold singularities are blown-up, the orientifolds will no longer intersect \cite{Junghans:2023yue}.
In that case, evaluating the backreaction within a perturbative framework may become more feasible \cite{Junghans:2020acz, Emelin:2022cac}. 
Furthermore, we have only investigated the untwisted closed string sector.
It would be interesting to explore the open string sector as well as the twisted states after blowing-up the singularities. We leave these interesting questions for future work.

\section*{Acknowledgments}
We would like to thank Vincent Van Hemelryck for discussions and comments on the construction, 
as well as valuable feedback on the first draft. 
G.T. is partially supported by the NSF grant PHY-2210271 and by the Lehigh University CORE grant with grant ID COREAWD40.

\end{document}